# ACCOMMODATING SAMPLE SIZE EFFECT ON SIMILARITY MEASURES IN SPEAKER CLUSTERING

*Alexander Haubold and John R. Kender*

Department of Computer Science, Columbia University

## ABSTRACT

We investigate the symmetric Kullback-Leibler (KL2) distance in speaker clustering and its unreported effects for differently-sized feature matrices. Speaker data is represented as Mel Frequency Cepstral Coefficient (MFCC) vectors, and features are compared using the KL2 metric to form clusters of speech segments for each speaker. We make two observations with respect to clustering based on KL2: 1.) The accuracy of clustering is strongly dependent on the absolute lengths of the speech segments and their extracted feature vectors. 2.) The accuracy of the similarity measure strongly degrades with the length of the shorter of the two speech segments. These effects of length can be attributed to the measure of covariance used in KL2. We demonstrate an empirical correction of this sample-size effect that increases clustering accuracy. We draw parallels to two Vector Quantization-based (VQ) similarity measures, one which exhibits an equivalent effect of sample size, and the second being less influenced by it.

*Index Terms*— Clustering methods, Speech analysis

## 1. INTRODUCTION

Automatic segmentation, clustering, and labeling of speaker data is a vital part of multimedia indexing, and is required for determining structure in video contents. This topic has found widespread attention in broadcast news analysis for speaker identification and tracking, only a few of which are mentioned here [5,8,9]. In the domain of instructional videos some work is available for discussion scene analysis [7].

Speaker clustering is based on the comparison of features extracted from a speaker segmented audio stream. Rarely are individual segments equal in length, and extracted features such as n-dimensional MFCCs and LPCs are computed on small fixed-length frames with constant sampling periods. Speaker segments of different length are then represented by a varying number of n-dimensional feature vectors. While distance metrics like the KL2 allow for such variation, side effects occur.

We show through simulation and empirical evidence that the KL2 distance between short audio segments (< 5 seconds) exhibits large degradation in performance. We also show that a similar effect exists for comparisons between differently-sized feature segments. We have observed decreasing accuracy the larger the difference in length of two audio segments.

We evaluate two alternative speaker clustering approaches based on vector quantization. The method proposed in [10] as a closed-form solution to KL2 for image comparison exhibits a similar trend in sample size effect. Smaller sets of features generally result in less accurate comparisons. A modified version of VQ for speaker identification [6] explicitly takes into consideration a correcting factor for sample size. Comparisons based on this measure show less biased results with respect to the size of a speaker's feature set.

## 2. BACKGROUND

The KL2 distance is used frequently in the context of speaker clustering based on MFCC features. It is not always clear from related literature under what conditions clustering took place. In [9] speaker data from news broadcasts is clustered using KL2, but it is not clear whether speech segments are equal or highly similar in size. In the news domain, however, speech segments tend to be short, given frequent cuts to other anchors and reporters. The authors of [5] argue that the Bayesian Inference Criterion (BIC) and other distance metric based methods often suffer in estimation error due to insufficient data. They observed degraded performance for segments of very short duration, e.g. 2 seconds, and attribute this phenomenon to insufficient data in the estimation of the covariance. Work in [8] presents a modified KL2 distance measure, in which the mean component is removed. The authors argue that the mean is easily biased due to various environment conditions. Most (80%) of their speaker segments are between 3 and 15 seconds in length. They have empirically observed that very short segments (< 2-3 seconds in length) dramatically decrease performance of segmentation and tracking. Work in [7] uses the KL2 distance for clustering speaker segments in the domain of instructional video. It is unclear what size distribution their speech data has, although we can assume significant differences in segment lengths given the genre of video.

## 3. SPEAKER CLUSTERING USING KL2

### 3.1. KL2 Clustering Approach

The symmetric KL distance is defined for two given random variables A and B with Gaussian distributions as:

$$KL2(A,B) = C(A,B) + M(A,B)$$

$$C(A,B) = \tfrac{1}{2}\left[tr(\sigma_A^{-1}\sigma_B) + tr(\sigma_B^{-1}\sigma_A)\right] - d \quad (1)$$

$$M(A,B) = (\mu_A - \mu_B)(\sigma_A^{-1} + \sigma_B^{-1})(\mu_A - \mu_B)^T$$

where $\sigma$ is the covariance matrix, $\mu$ the mean vector, and $d$ is the dimensionality of the feature vector. A typical value for dimensionality $d$ is 13 for MFCC features, representing the energy coefficient $d(0)$, and the 12 first MFCC coefficients $d(1) - d(12)$.

Based directly on its derivation in [1], the KL2 function is comprised of two parts: a portion, $C(A,B)$, which is strictly computed from the covariance and a portion, $M(A,B)$, which also includes the mean vectors of the feature set. $M(A,B)$ often directly reflects, among other physical influences, possible environmental changes in the audio source, e.g., volume shifts. Some work suggests that results improve when this term is not considered at all [7].

### 3.2. KL2 Simulation

Since KL2 is a distance measure, feature sets with similarities produce a KL2 value towards 0, while segments with differences produce KL2 values > 0. We observe significant differences in the KL2 measure for feature sets from one speaker in which at least one segment is short. And, during clustering, long audio segments (> 20 seconds) also are observed to converge to clusters faster and with more accuracy than shorter segments. Therefore, we set up two simulations to test the KL2 distance for these observed length effects. In *Experiment 1*, random datasets of variable-sized MFCC feature sets are generated, and their simulated KL2 distances are determined in a matrix of (*segment A length*) × (*segment B length*). In *Experiment 2*, MFCC features from a single speaker are used to create new variable-sized feature sets, and their KL2 distances are computed in a similar matrix. The first experiment uses simulated data to illustrate the general trend in KL2 over differently sized speech segments (Figure 1), while the second experiment validates this trend on real data (Figure 2). Trends from the simulation are supported by real data.

### 3.3. Observations

The KL2 measure for comparisons between differently sized speech segments shows significantly degraded results depending on the length of the shorter segment. Additionally, smaller feature sets have obvious disadvantages over larger ones. A closer analysis of the KL2 metric, which outlines the source of the length effect, is presented in Figure 5 and 6. The second term of $C(A,B)$, itself an asymmetric KL distance $tr(\sigma_B^{-1}\sigma_A)$ for speech segments A and B, responds to the length of B strongly in what appears to be an inverse power fashion (Figure 5); variations due to the length of A are more moderate (Figure 6). The symmetric KL2 distance therefore includes two such asymmetric terms, which behave similarly (Figure 4).

The reason behind the degraded KL2 distances for short speech segments is the lack of sufficient samples that can be used to accurately model the parameters of the Gaussian distributions in the d-dimensional feature space. These observations suggest that, contrary to what has been suggested in the literature, using short fixed-length speech segments for comparing and clustering can be highly inaccurate, even if only one segment is short. Many media, such as news or presentation videos, utilize such short segments, and although identifiable by humans, machine clustering often fails.

### 3.4. Empirical Solution

The KL2 distance measure is derived under two critical simplifying assumptions [1]: first, that the MFCC vectors are distributed as a d-dimensional Gaussian, and, second, that the sample means and covariances are perfect estimators of the population means and covariances. The first assumption is necessary to allow a closed form evaluation of a d-dimensional integral. The second assumption eliminates the need to model the effects of the two samples' lengths on the standard errors of their statistics: the samples are assumed to be infinitely long with zero standard error. A Gaussian mixture model would be more appropriate, but deriving a closed form KL2 for it presents formidable analytic difficulties. Similarly, in practice, the estimated statistics show increasing error as sample length diminishes. However, the analytic modeling of the impact of standard errors also is challenging, even under the assumption of a single d-dimensional Gaussian. The KL2 measure, then, has likely been used without an understanding of these substantial limitations.

In lieu of an analytic closed-form computation parameterized by the incoming lengths, we investigated a possible empirical solution, which adjusts the KL2 distance, based on a simulated model's response with identical distributions:

$$KL2'(A,B) = \frac{KL2(A,B)}{KL2sim(|A|,|B|)} \quad (2)$$

where $KL2'(A,B)$ is the adjusted KL2 distance for speech segment pair $(A,B)$ with lengths $(|A|,|B|)$, and $KL2sim(|A|,|B|)$ is the simulated KL2 value taken from the results of Experiment 1. Results for this solution are presented in Figure 2. With the exception of a few outliers,

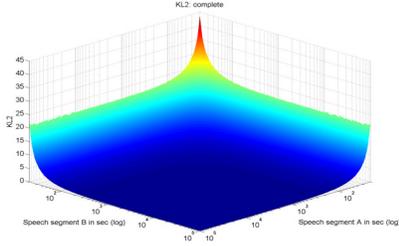
Figure 1: KL2 distance length effect. Simulation on randomly generated normal distributions: Comparing two small-sized feature vector sets (< 200 vectors) introduces estimation errors due to limited sampling size.

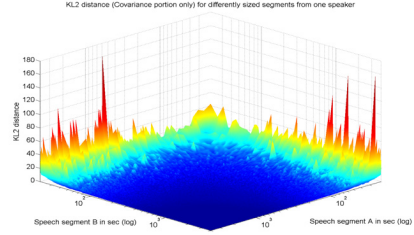
Figure 2: Real speech data. Random subsets (1 to 252 seconds) are extracted from one speaker's feature set, and compared pair-wise. Comparisons involving short segments (either one or both) result in higher KL2 distance values, and thus less similarity.

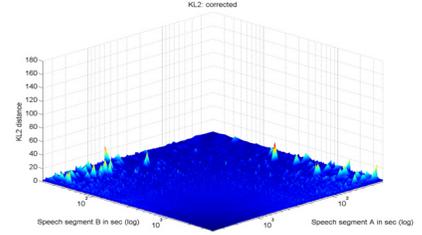
Figure 3: KL2' responses for real speech data, compensating for length effect. The response of KL2' more uniformly reflects distance. Note that vertical axis is 4 times that of Figure 3 and Figure 4.

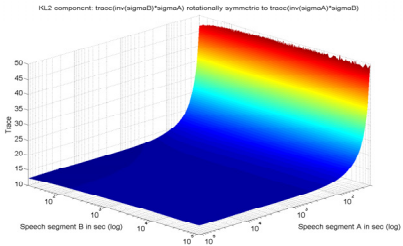
Figure 4: Term $tr(\sigma_B^{-1}\ \sigma_A)$ evaluated for variable feature sets A and B. Graph is rotationally symmetric to first term, $tr(\sigma_A^{-1}\ \sigma_B)$.

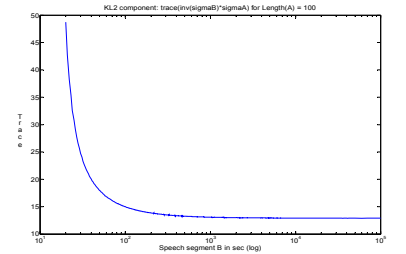
Figure 5: Graphical analysis of KL2 covariance term $tr(\sigma_B^{-1}\ \sigma_A)$, evaluated at fixed feature set A length=100, and variable feature set B length; showing a factor of 3 variation.

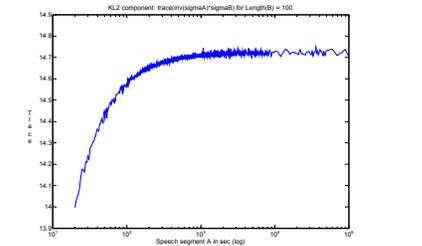
Figure 6: Term $tr(\sigma_B^{-1}\ \sigma_A)$ evaluated at fixed feature set B length=100, and variable feature set A length. Vertical axis exaggerates the response variation, which is within 5%.

the KL2' distance for short feature sets is more comparable with that of long ones for data from one speaker.

### 3.5. Results

We have tested our solution for clustering on data sets containing 5 and 29 speakers with significantly different-sized speech (6-189 seconds). Figure 7 outlines the data set for one presentation video with 6 presentations. Results for our empirical solution are very favorable and include two observations. Firstly, after applying the correction factor, short speaker segments are united with the true cluster to which they belong, thus both validating and correcting the previous shortcomings of the KL2 distance. Secondly, the threshold values between true and false clusters in the dendrogram representation are more apparent due to the consideration of the KL2 length effect. Previously, longer speaker segments clustered more easily together, an effect ascribed mainly to their length. With the inclusion of the length effect factor, it is possible for shorter segments to correctly cluster earlier with longer ones, if they belong to the same speaker.

Clustering results improved locally on a presentation level (not shown), as well as globally over a series of presentations. We have tested the KL2' correction with the original KL2 distance, which included covariance and mean terms, as well the modified KL2 distance without the mean proposed in [7]. Results improved for both, and we note that KL2' without the mean term performed slightly better in clustering.

### 4. SPEAKER CLUSTERING USING VQ

Vector quantization is derived as a closed-form expression to KL2, which does not have a closed-form solution [10]. The clustering approach based on VQ density estimates exhibits a similar degradation of results for comparisons of speech segments with highly varying lengths from the same speaker. We have performed a simulation similar to that for KL2, and compared the clustering performance between the original and the corrected VQ similarity measure. Results (not shown) are again favorable towards a correcting measure. Figure 8 highlights the observed distortion in sample sizes ([0.5 .. 0.8]), which is less pronounced in its absolute value than sample size on clustering performance, and present the that of KL2 ([0 .. 45]), but still high enough to cause clustering of segments with highly varying lengths to fail.

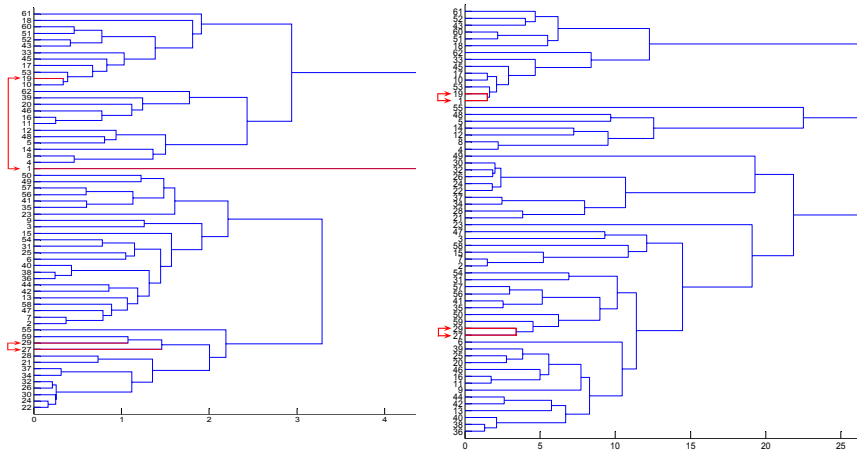
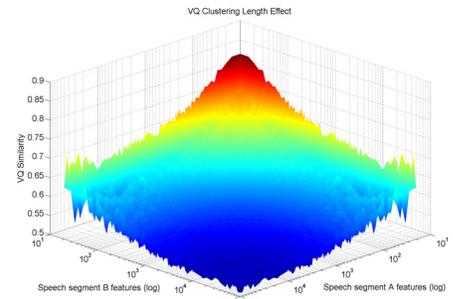

Figure 8: VQ distance length effect.

Figure 7: Results of speaker clustering with and without correcting for length. Shown is a dendrogram excerpt for one presentation video. Highlighted clusters are the results of comparing between segments of significantly different length. With original KL2 (left), segments of the same speaker but of different lengths do not cluster well. Once adjusted with an empirical correction factor (right), clustering improves. Correct clusters: (top marked) ids={1,19} (6 and 57 secs), (bottom marked) ids={27,29} (17 and 59 secs).

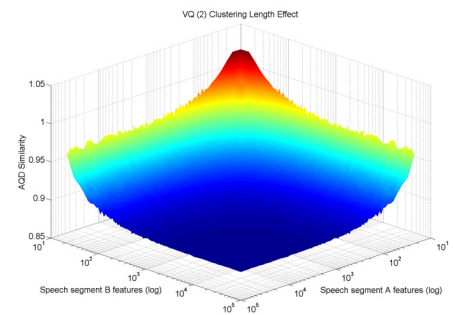

Figure 9: Modified VQ distance length effect.

The approach presented in [6] aims to optimize VQ-based speaker identification. The featured algorithm addresses the sample size effect in its computation of the match score which produces the average quantization distortion (AQD). This step divides the cumulative matching score by the number of speaker feature samples.

We have performed a simulation to describe the effect of sample size on clustering performance, and present the similarity matrix in Figure 9. While a similar trend to that of KL2 and VQ appears, the distortion along the y-axis is not pronounced ([0.89 .. 1.02]), and does not affect clustering results as heavily. From speaker clustering experiments, we have determined that an additional correction does not improve results.

## 5. CONCLUSION

We have noted sensitivities of the KL2 distance and VQ-based density estimates to the length of their input feature sets, which results in deviations between comparisons of short but practical speech segments. We have presented a solution based on empirical evidence for both approaches. Future investigations include a more thorough analysis of the MFCC input feature set distributions, and other non-analytic accommodations of short non-Gaussian speech.